\documentclass[a4paper,12pt]{article}
\usepackage{jheppub}

\usepackage{amsmath}
\usepackage{amssymb}
\usepackage{multirow}
\usepackage{graphicx}

\newcommand{\abs}[1]{\lvert #1 \rvert}

\numberwithin{equation}{section}

\def\be{\begin{equation}}
\def\ee{\end{equation}}

\title{The Continuous Orbifold of $\boldsymbol{{\cal N}=2}$ \\ Minimal Model Holography}

\author{Matthias R.\ Gaberdiel}
\author{and Maximilian Kelm}

\affiliation{Institut f\"ur Theoretische Physik, ETH Z\"urich, \\
CH-8093 Z\"urich, Switzerland}

\emailAdd{gaberdiel@itp.phys.ethz.ch}
\emailAdd{mkelm@itp.phys.ethz.ch}

\abstract{For the 
$\mathcal{N}=2$ Kazama-Suzuki models
that appear in the duality with a higher spin theory
on AdS$_3$ it is shown that the large level limit 
can be interpreted as a continuous orbifold
of $2N$ free bosons and fermions by the group 
$\mathrm{U}(N)$. In particular, we show that the subset of coset representations
that correspond to the perturbative higher spin degrees of freedom are precisely
described by the untwisted sector of this ${\rm U}(N)$ orbifold. 
We furthermore identify the twisted sector ground states of the orbifold with
specific coset representations, and give various pieces of evidence in favour of 
this identification.}

\date{\today}

\begin{document}
\maketitle

\section{Introduction}
Dualities between Vasiliev higher spin theories on Anti-de Sitter spacetimes \cite{Vasiliev:2003ev}
and conformal field theories constitute a promising way towards understanding the AdS/CFT correspondence.
In particular, dualities of this type are `vector-like', and hence contain considerably fewer
degrees of freedom than the `adjoint-like' theories appearing in the stringy AdS/CFT duality. 
Furthermore, they are in a sense weak-weak dualities, and  may therefore be amenable to
a perturbative proof (see  \cite{Giombi:2012ms} and  \cite{Gaberdiel:2012uj} for 
reviews). This could then form the seed towards establishing the full AdS/CFT
correspondence, at least at the tensionless point where a description in terms of a higher spin theory
is expected to arise.

In order to understand the connection between higher spin theories and string theory in more
detail, it is useful to study supersymmetric variants of the higher spin CFT duality.
Furthermore, it is natural to try to do so first in a low-dimensional setting where 
the higher spin theories are considerably simpler \cite{Prokushkin:1998bq,Prokushkin:1998vn},
and also much is known about the stringy AdS/CFT duality, see \cite{David:2002wn} for a review.
With this in mind, a duality between a family of $\mathcal{N}=4$ supersymmetric coset CFTs in $2$d, and a 
supersymmetric higher spin theory on AdS$_3$
was  proposed in \cite{Gaberdiel:2013vva} and subsequently tested and extended in various ways
\cite{Creutzig:2013tja,Candu:2013fta,Gaberdiel:2014yla,Beccaria:2014jra,Creutzig:2014ula}.
This duality is 
a natural supersymmetric generalisation of the original bosonic higher spin CFT duality of 
\cite{Gaberdiel:2010pz}.

A particularly interesting limit of the ${\cal N}=4$ cosets arises for the case when one 
of the levels is taken to infinity since one may then hope to make contact with
the D1-D5 system; this will be explored  
in detail \cite{GGprep}.  As a preparation for this analysis, we study in this paper the
large level limit of the ${\cal N}=2$ Kazama-Suzuki cosets  \cite{Kazama:1988qp,Kazama:1988uz}
that occur in the higher spin duality with an ${\cal N}=2$ supersymmetric higher spin
theory on AdS$_3$ \cite{Creutzig:2011fe,Candu:2012jq}.
More concretely, we consider the cosets 
\be\label{eq:coset_KS}
 \frac{\mathfrak{su}(N+1)_{k}\oplus \mathfrak{so}(2N)_1}{\mathfrak{su}(N)_{k+1} 
 \oplus \mathfrak{u}(1)_{N(N+1)(N+k+1)}}  
\ee
in the limit where the level $k$ is taken to infinity. We give convincing evidence that the limit theory
has an interpretation as a ${\rm U}(N)$ orbifold of  $2N$ free fermions and bosons that both
transform as  ${\bf N}\oplus \bar{\bf N}$  under ${\rm U}(N)$.\footnote{The idea
that the limit theory has such an interpretation was already mentioned in \cite{Restuccia:2013tba}, 
following on from the analysis of \cite{Fredenhagen:2012bw},
where this was shown explicitly for $N=1$.}
This is the natural generalisation of the
bosonic analysis of \cite{Gaberdiel:2011aa}, where it was shown that the cosets 
\be\label{eq:coset_WN}
\frac{\mathfrak{su}(N)_k\oplus\mathfrak{su}(N)_1}{\mathfrak{su}(N)_{k+1}}
\ee
admit a description in terms of an orbifold of $N-1$ free bosons by the Lie group $\mathrm{SU}(N)$.
In each of these
cases the limit is taken in the spirit of \cite{Runkel:2001ng} (rather than say \cite{Roggenkamp:2003qp}), 
see also \cite{Fredenhagen:2010zh,Fredenhagen:2012rb} for other instances (closely 
related to the topic of this paper) where this kind of 
construction has  been considered. 
We should also mention that this orbifold picture is the natural $2$d analogue of the ${\rm U}(N)$ 
(or ${\rm O}(N)$) singlet sector of a theory of free bosons or fermions in $3$d that played 
an important role in the higher spin  CFT duality in one dimension higher 
\cite{Klebanov:2002ja,Sezgin:2003pt}. 

We shall consider the usual charge conjugation modular invariant of the coset (\ref{eq:coset_KS})
whose duality to the higher spin theory on AdS$_3$ was explored in  
\cite{Creutzig:2011fe,Candu:2012jq,Henneaux:2012ny,Hanaki:2012yf,Ahn:2012fz,Candu:2012tr}.
In particular, we shall see that the part of the CFT spectrum that corresponds to the perturbative
higher spin degrees of freedom 
\be\label{pert}
{\cal H}_{\rm pert} = \bigoplus_{\Lambda} {\cal H}_{(0;\Lambda)} \otimes \overline{\cal H}_{(0;\Lambda^\ast)}
\ee
can be identified, for $k\rightarrow \infty$,  with the subspace of the free field theory of  $2N$ bosons and fermions 
that are  singlets  with respect to ${\rm U}(N)$, i.e., with the untwisted sector of the continuous orbifold.
The remaining coset primaries,
i.e., those of the form $(\Lambda_+;\Lambda_-)$ with $\Lambda_+\neq 0$, can then be interpreted
in terms of the various twisted sectors of the continuous orbifold. In fact, as is familiar from usual orbifolds,
the untwisted sector is not modular invariant by itself, and the twisted sectors are required in order
to restore modular invariance. For the case at hand where we have supersymmetry, the identification 
of the different coset primaries with the twisted sectors can be worked out in detail, and a number of 
non-trivial consistency checks can be performed. In particular, we have compared the conformal 
dimension of the twisted sector ground states with that calculated from the coset viewpoint; we have also
determined the fermionic excitation spectrum directly from the coset perspective. 
\medskip

The paper is organised as follows.
In section~\ref{sec:2} we introduce our conventions and review briefly the  relevant
Kazama-Suzuki coset models as well as some of their low-lying 
representations. In section~\ref{sec:3} we identify the subsector of perturbative states (\ref{pert})
with the untwisted sector of the continuous orbifold, i.e., with the subspace of the free
field theory that consists of the singlets under the ${\rm U}(N)$ action. In particular,
we show in sections~\ref{sec:3.1} and \ref{sec:3.2} that the partition functions of the two descriptions agree.
Section~\ref{sec:twist} is dedicated to the analysis of the twisted sectors. We identify all twisted sector ground 
states with coset primaries, see eq.~(\ref{eq:gs0}), and show that the conformal dimensions
agree. Furthermore, we compare their fermionic excitation spectrum (see section~\ref{sec:ex}),
as well as the structure of their BPS descendants (see section~\ref{sec:BPS}), and find beautiful agreement.
Section~\ref{sec:5} contains our conclusions as well as a brief outlook. We have relegated some 
background information and a few detailed computations to three appendices.

\section{The \texorpdfstring{$\boldsymbol{\mathcal{N}=2}$}{N=2} Kazama-Suzuki coset model}\label{sec:2}

Let us begin by introducing our conventions for the ${\cal N}=2$ superconformal field theories
that appear in the 
duality to the ${\cal N}=2$ supersymmetric higher spin theory on AdS$_3$ \cite{Creutzig:2011fe}. 
The relevant  cosets \cite{Kazama:1988qp,Kazama:1988uz} are (see \cite{Candu:2012jq} for our conventions)
\be
\frac{\mathfrak{su}(N+1)^{(1)}_{k+N+1}}{\mathfrak{su}(N)^{(1)}_{k+N+1} \oplus \mathfrak{u}(1)^{(1)}_\kappa} 
\ \cong \ \frac{\mathfrak{su}(N+1)_{k}\oplus \mathfrak{so}(2N)_1}{\mathfrak{su}(N)_{k+1} \oplus \mathfrak{u}(1)_\kappa}  \ ,
\ee
where the second description is in terms of the bosonic affine algebras.
Here the level of the $\mathfrak{u}(1)$ factor equals $\kappa=N(N+1)(N+k+1)$, and the central charge is
\be
c = (N-1) + \frac{kN(N+2)}{k+N+1} - \frac{(k+1)(N^2-1)}{k+N+1} = \frac{3kN}{k+N+1} \ .
\ee
The subgroup of the denominator $\mathrm{SU}(N)\times \mathrm{U}(1)$ is `embedded' into 
$\mathrm{SU}(N+1)$ via the ($N$-to-one) mapping
\be \label{suem}
(v,w) \mapsto  \begin{pmatrix}
\bar{w}v&0\\
0& w^N \end{pmatrix}\ , 
\ee
where $w \in {\rm U}(1)$ is a phase, while $v\in {\rm SU}(N)$ is an $N\times N$ matrix. Similarly, the `embedding'
into $\mathrm{SO}(N,N)$ (whose complexified Lie algebra agrees with the complexification of $\mathfrak{so}(2N)$)
is defined by
\be \label{soem}
(v,w) \mapsto  \begin{pmatrix}
\bar{w}^{N+1}v & 0\\ 
0& w^{N+1}\bar{v}
\end{pmatrix}\ ,
\ee
see \cite{Candu:2012jq} for more details. 
Our conventions are chosen so that the free fermions and bosons have $\mathrm{U}(1)$ charge $\pm(N+1)$.

The representations of the coset are labelled by $(\Lambda_+;\Lambda_-,\ell)$, where $\Lambda_+$ 
is an integrable  weight of $\mathfrak{su}(N+1)_k$, $\Lambda_-$ an integrable weight of 
$\mathfrak{su}(N)_{k+1}$, while  $\ell$ denotes the $\mathfrak{u}(1)$ charge.\footnote{Strictly
speaking, the coset representations are also labelled by the representation
of $\mathfrak{so}(2N)_1$. In this paper we shall only consider the NS sector of the free fermions, i.e., 
we shall take the $\mathfrak{so}(2N)_1$ representation to be either the vacuum or the vector representation.} 
The selection rule is 
\be\label{sel}
\frac{|\Lambda_+|}{N+1} - \frac{|\Lambda_-|}{N} - \frac{\ell}{N(N+1)} \in \mathbb{Z} \ ,
\ee
where $|\Lambda| = \sum_{j} j \Lambda_j$, 
and we have the field identification
\be\label{N2fi}
(\Lambda_+;\Lambda_{-},\ell) \ \cong \ \bigl(J^{(N+1)}\, \Lambda_+; J^{(N)}\, \Lambda_{-}, \ell-(k+N+1) \bigr) \ ,
\ee
where $J$ denotes the usual outer automorphism, i.e., it maps (for the case of $\mathfrak{su}(N+1)$)
\be
\Lambda = [\Lambda_0;\Lambda_1,\ldots,\Lambda_N] \ \mapsto  \ 
J^{(N+1)}\, \Lambda = [\Lambda_N; \Lambda_0,\Lambda_1,\ldots,\Lambda_{N-1}] \ . 
\ee
Since the field identification acts simultaneously on a weight in $\mathfrak{su}(N+1)$ 
and $\mathfrak{su}(N)$, it has order $N(N+1)$; this then ties together with the fact that 
the $\mathfrak{u}(1)$ charge $\ell$ is defined modulo $\kappa=N(N+1)(N+k+1)$. 
\smallskip

\noindent 
The conformal dimension of the representation $(\Lambda_+;\Lambda_{-},\ell)$ equals
\be\label{hdef}
h(\Lambda_+;\Lambda_{-},\ell) = \frac{C^{(N+1)}(\Lambda_+)}{N+k+1} 
- \frac{C^{(N)}(\Lambda_{-})}{N+k+1} - \frac{\ell^2}{2N(N+1)(N+k+1)} + n \ , 
\ee
where $n$ is a half-integer, describing the `level' at which $(\Lambda_{-},\ell)$ appears in the representation 
$\Lambda_{+}$, and $C^{(N)}(\Lambda)$ is the quadratic Casimir
of the $\mathfrak{su}(N)$ weight $\Lambda$. Finally, the ${\rm U}(1)$ charge (with respect to the ${\rm U}(1)$ generator of the 
superconformal ${\cal N}=2$
algebra) equals
\be\label{qdef}
q(\Lambda_+;\Lambda_{-},\ell) = \frac{\ell}{N+k+1} + s \ , 
\ee
where $s\in\mathbb{Z}$ denotes the charge contribution of the descendants. For example, the representation
\be
({\rm f};0,N) : \qquad h = \frac{N}{2(N+k+1)} \ , \quad q = \frac{N}{N+k+1} \ , 
\ee
where ${\rm f}$ denotes the fundamental representation of $\mathfrak{su}(N+1)$, 
describes a chiral primary, as does 
\be\label{0fh}
h(0;{\rm f},-(N+1))  = \frac{1}{2} - \frac{(N^2-1)}{2N(N+k+1)} - \frac{(N+1)}{2N(N+k+1)} = 
\frac{k}{2(N+k+1)} \ , 
\ee
for which the ${\rm U}(1)$ charge equals
\be\label{0fq}
q(0;{\rm f},-(N+1))  = \frac{-(N+1)}{N+k+1} +1  = \frac{k}{N+k+1}\ .
\ee
Here the additional terms in (\ref{0fh}) and (\ref{0fq}) appear because 
for $(0;{\rm f},-(N+1))$ the representation of the denominator arises only at the first excited level.

\section{The continuous orbifold: the untwisted sector}\label{sec:3}

We are interested in taking the $k\rightarrow \infty$ limit of these cosets. For the case 
$N=1$ with $c=3$, this was worked out in some detail in \cite{Fredenhagen:2012bw}, where
it was shown that the resulting theory can be interpreted in terms of a continuous ${\rm U}(1)$
orbifold. Here we want to extend the discussion to general $N$. The idea that the limit  theory
may be interpreted in terms of a ${\rm U}(N)$ orbifold was already sketched in \cite{Restuccia:2013tba}; 
in the following, we shall pursue a somewhat different approach and be much more explicit.

The discussion of \cite{Candu:2012jq} as well as the analogous analysis in  \cite{Gaberdiel:2011aa} 
suggests that the underlying free theory consists of $2N$ free bosons and free fermions that transform as
\be\label{transf}
{\bf N}_{-(N+1)} \oplus \bar{\bf N}_{N+1} 
\ee
with respect to $\mathfrak{su}(N) \oplus \mathfrak{u}(1)$ in the denominator. The relevant orbifold group is therefore
${\rm SU}(N)\times {\rm U}(1)$, or equivalently $\mathrm{U}(N)$,\footnote{The 
discrete subgroup of 
${\rm SU}(N)\times {\rm U}(1)$ that needs to be factored out to obtain $\mathrm{U}(N)$ acts trivially.} 
where the group acts simultaneously on both left- and right-movers. 
\smallskip

\noindent One reason in favour of this idea is that the central charge approximates in this limit 
\be
c = \frac{3 kN}{k+N+1} \cong 3 N \ ,
\ee
in agreement with a description in terms of $2N$ free bosons and fermions. Furthermore,
the ground states of the representations $(0;{\rm f}, -(N+1))$ and $(0;\bar{\rm f}, (N+1))$ can be identified with
the ${\bf N} + \bar{\bf N}$ free fermions since their conformal dimension and $\mathfrak{u}(1)$ charge 
become in this limit 
\be
h (0;{\rm f}, -(N+1)) = h (0;\bar{\rm f}, (N+1)) = \frac{1}{2} \ , 
\ee
as well as 
\be
q  (0;{\rm f}, -(N+1)) = + 1 \ , \qquad  q (0;\bar{\rm f}, (N+1)) = - 1 \ .
\ee
Each of these representations has two ${\cal N}=2$ descendants with $h=1$, which can in turn 
be identified with the free bosons. For the actual coset partition function, left- and right-movers are
grouped together, i.e., $(0;{\rm f}, -(N+1))$ for the left-movers appears together with 
$(0;\bar{\rm f}, (N+1))$ for the right-movers, etc., and this is precisely what the ${\rm U}(N)$ singlet
condition achieves.
\medskip

Concretely, we therefore claim that the untwisted sector of the $\mathrm{U}(N)$ orbifold
of $2N$ free bosons and fermions, transforming as in (\ref{transf}), 
corresponds to the subsector of the coset theory
\be\label{Hun}
\mathcal{H}_0 = \bigoplus_{\Lambda,u} \, \mathcal{H}_{(0;\Lambda,u)}
\otimes \bar{\mathcal{H}}_{(0;\Lambda^{\ast},-u)}
\ee
in the limit $k \to \infty$. Here the sum runs over all representations $\Lambda$ that appear in
finite tensor powers of the fundamental or anti-fundamental representation of $\mathfrak{su}(N)$ --- in the
limit $k\rightarrow\infty$, the $k$-dependent bound on the integrable  $\mathfrak{su}(N)_{k+1}$ representations 
disappears --- and $\Lambda^\ast$ denotes the representation conjugate to $\Lambda$. Furthermore,
$u$  must satisfy the selection rule that $(N+1) |\Lambda| + u = 0$ mod $N(N+1)$. 

In the following we will give strong evidence in favour of this claim by showing that the partition functions
agree. In section~\ref{sec:twist} we shall then also explain how the twisted sectors of the continuous
orbifold can be understood from the coset viewpoint.

\subsection{The partition function from the coset}\label{sec:3.1}

We want to show that the spectrum of the untwisted sector of the $\mathrm{U}(N)$ orbifold
coincides with eq.~(\ref{Hun}) by comparing partition functions. In order to do so, we need to understand
the character of the coset representations $(0;\Lambda,u)$ in the limit $k \to \infty$.
For large $k$, the character of an affine representation $\Lambda$ of $\mathfrak{su}(N)_k$ is given by
\be 
\mathrm{ch}^{N,k}_{\Lambda}(v;q) = \frac{q^{h_{\Lambda}^{N,k}}\bigl[\mathrm{ch}^N_{\Lambda}(v)
+\mathcal{O}(q^{k-\sum_i\Lambda_i+1})\bigr]}
{\prod_{n=1}^{\infty}\bigl[(1-q^n)^{N-1}\prod_{i \neq j}(1-v_i \bar{v}_j q^n)\bigr]}\ .
\ee
Here $v_i$ are the eigenvalues of $v \in \mathrm{SU}(N)$, $\mathrm{ch}^N_{\Lambda}(v)$
is the character of $\Lambda$ restricted to the zero mode subalgebra $\mathfrak{su}(N)$,
$\Lambda_i$ are the Dynkin labels of $\Lambda$,
and we define
\be
 h_{\Lambda}^{N,k}=\frac{C^{(N)}(\Lambda)}{N+k}\ ,
\ee
where $C^{(N)}(\Lambda)$ is, as before, the quadratic Casimir of $\Lambda$.
For example, the vacuum character $\mathrm{ch}^{N+1,k}_0(v,w;q)$ of $\mathfrak{su}(N+1)_k$ with 
$v \in \mathrm{SU}(N)$ and $w \in \mathrm{U}(1)$ embedded into $\mathrm{SU}(N+1)$
as in (\ref{suem})  equals
\be 
\mathrm{ch}^{N+1,k}_0 = 
 \frac{1+\mathcal{O}(q^{k+1})}{\prod_{n=1}^{\infty}\Bigl[(1-q^n)^{N} 
\prod_{i \neq j}(1-v_i \bar{v}_j q^n)\prod_{i=1}^N\bigl[(1-\bar{w}^{N+1}v_iq^n)(1-w^{N+1}\bar{v}_iq^n)\bigr]\Bigr]}\ .
\ee
Moreover, the representations of the $\mathfrak{so}(2N)_1$ factor in the numerator are the vacuum and vector
representation, as well as  either of the two spinor representations. In terms of the free fermions 
(that are equivalent to $\mathfrak{so}(2N)_1$), the former two correspond to the NS sector, while the latter 
are accounted for in terms of the R sector. In the following we shall concentrate on the 
NS sector\footnote{In the duality to the higher spin theory on AdS$_3$ only the NS-NS sector
plays a role since the conformal dimension of the RR sector states is proportional to the central charge, see
the discussion in \cite{Gaberdiel:2011nt}.} for which the contribution of the  $2N$ free fermions equals
\be
 \theta(v,w;q)=\prod_{n=1}^\infty \prod_{i=1}^N (1+\bar{w}^{N+1}v_iq^{n-\frac{1}{2}})
 (1+w^{N+1}\bar{v}_iq^{n-\frac{1}{2}})\ .
\ee
The characters of the denominator, on the other hand,
are given in that limit by
\be \mathrm{ch}^{N,k+1}_{\Lambda,u}(v,w;q)
= \frac{q^{h_{\Lambda}^{N,k+1}+\frac{u^2}{2\kappa}} \bigl(w^u +\mathcal{O}(q^{\frac{\kappa}{2}-|u|})\bigr)
\bigl(\mathrm{ch}_{\Lambda}^N(v)+\mathcal{O}(q^{k-\sum_i\Lambda_i+2})\bigr)}
{\prod_{n=1}^{\infty}\bigl[(1-q^n)^{N}\prod_{i \neq j}(1-v_i \bar{v}_j q^n)\bigr] }\ .
\ee
The coset character associated to $(0;\Lambda,u)$ is then given by the branching function
$b_{0;\Lambda,u}^{N,k}(q)$, which is defined by
\be
\mathrm{ch}^{N+1,k}_0(v,w;q)\,\theta(v,w;q)=\sum_{\Lambda,u} 
b_{0;\Lambda,u}^{N,k}(q)\, \mathrm{ch}^{N,k+1}_{\Lambda,u}(v,w;q)\ .
\ee
Combining the explicit expressions given above,
the branching functions take the form (see also \cite{Candu:2012jq})
\begin{equation}
\label{eq:branching}
 b_{0;\Lambda,u}^{N,k}(q) = q^{-h_{\Lambda}^{N,k+1}-\frac{u^2}{2\kappa}}\Bigl[a_{0;\Lambda,u}^N(q)
+ \mathcal{O}(q^{k-\sum_i\Lambda_i+2})+\mathcal{O}(q^{\frac{\kappa}{2}-|u|})\Bigr]\,,
\end{equation}
where $a_{0;\Lambda,u}^N(q)$ is the multiplicity of $w^u \mathrm{ch}_{\Lambda}^N(v)$ in
\begin{equation}
\label{eq:coset_pfnum}
\sum_{\Lambda,u} \, a_{0;\Lambda,u}^N(q) \, w^u \mathrm{ch}_{\Lambda}^N(v) = 
\prod_{n=1}^\infty \prod_{i=1}^N \frac{(1+\bar{w}^{N+1}v_iq^{n-\frac{1}{2}})
 (1+w^{N+1}\bar{v}_iq^{n-\frac{1}{2}})}{(1-\bar{w}^{N+1}v_iq^{n})
 (1-w^{N+1}\bar{v}_iq^{n})}\ .
\end{equation}
It therefore follows that the partition function $\mathcal{Z}_0$ of (\ref{Hun}) equals for $k\to\infty$ 
\begin{equation}
\label{eq:coset_pf0}
 \mathcal{Z}_0 = \lim_{k\to\infty} (q\bar{q})^{-\frac{c}{24}}\sum_{\Lambda,u} |b_{0;\Lambda,u}^{N,k}(q)|^2
= (q\bar{q})^{-\frac{N}{8}}\sum_{\Lambda,u}| a_{0;\Lambda,u}^N(q) |^2\ ,
\end{equation}
where we sum over all finite Young diagrams $\Lambda$ of at least $N-1$ rows, and $u$ must be of the form
$u=(N+1)(-|\Lambda|+n N)$ with $n \in \mathbb{Z}$.  In the second equality, we have used that
since $\Lambda$ and $u$ are finite (and do not grow with $k$), the prefactor in eq.~(\ref{eq:branching}),
$h_{\Lambda}^{N,k+1}+\frac{u^2}{2\kappa}$, vanishes in the limit, and the higher order terms 
in the bracket become irrelevant.

\subsection{Comparison with the untwisted orbifold sector}\label{sec:3.2}

We shall now compare this result to the ${\rm U}(N)$ orbifold of $2N$ 
free fermions and bosons
that transform as ${\bf N}\oplus \bar{\bf N}$ of $\mathrm{U}(N)$, cf., eq.~(\ref{transf}). 
Labelling again the elements of ${\rm U}(N)$ in terms of $\mathrm{SU}(N) \times \mathrm{U}(1)$ 
via the `embedding'
\be
\imath \colon (v,w) \mapsto w^{-(N+1)}\cdot v =\bar{w}^{(N+1)}\cdot v \ , 
\ee
the partition function with the insertion of these group elements takes the form
\begin{equation}
 \label{eq:orbfree}
 \imath(v,w)\cdot \mathcal{Z}_{\text{free}}
= (q\bar{q})^{-\frac{N}{8}}\prod_{n=1}^{\infty}\prod_{i=1}^N\frac{|1+\bar{w}^{(N+1)}v_iq^{n-\frac{1}{2}}|^2
|1+w^{N+1}\bar{v}_i q^{n-\frac{1}{2}}|^2}{|1-\bar{w}^{(N+1)}v_iq^n|^2
|1-w^{N+1}\bar{v}_i q^n|^2}\,,
\end{equation}
where we have used that the central charge equals $c=3N$.
The untwisted sector of this orbifold theory consists of the states that are
$\mathrm{U}(N)$ invariant. Put differently, the untwisted sector is therefore the 
multiplicity space of the trivial representation of $\mathrm{U}(N)$ acting on 
the free theory with partition function $\mathcal{Z}_{\text{free}}$.
Since \eqref{eq:orbfree} is, up to the prefactor, just the charge-conjugate square
of the coset numerator character \eqref{eq:coset_pfnum}, 
this amounts to finding the trivial representation in
\be
 (0;\Lambda_1,u_1) \otimes (0;\Lambda_2,u_2)
\ee
for some representations $(\Lambda_i,u_i)$ ($i=1,2$)
of $\mathfrak{su}(N)\oplus \mathfrak{u}(1)$, where the first factor
corresponds to the left-movers and the second one to the right-movers. 
This tensor product contains the trivial representation if and only if
$\Lambda_1=\Lambda_2^{\ast}$ and $u_1=-u_2$, where $\Lambda_2^{\ast}$ is
the representation conjugate to $\Lambda_2$, and it always does so with multiplicity one.
Thus we conclude that the partition function of the untwisted sector equals
\begin{equation}
 \mathcal{Z}_{\text{U}} =  (q\bar{q})^{-\frac{N}{8}}\sum_{\Lambda,u} |a_{0;\Lambda,u}^N(q) |^2\,,
\end{equation}
matching precisely \eqref{eq:coset_pf0}.
This yields convincing evidence that the coset subsector of states $(0;\Lambda,u)$
can indeed be described by the untwisted sector of the ${\rm U}(N)$ orbifold introduced above.

\section{Twisted sectors}\label{sec:twist}

The remaining states, i.e., those with $\Lambda_+\neq 0$, should then arise from the twisted sector of the
continuous orbifold.
In the following we shall be able to make this correspondence rather concrete. The main reason why
we can be much more explicit (see eq.~(\ref{eq:gs_num}) below)
than in the corresponding bosonic analysis of  \cite{Gaberdiel:2011aa}
is that the ${\cal N}=2$ superconformal symmetry is quite restrictive and in particular implies that the 
ground state energy of the twisted sectors is linear in the twist.
\smallskip

To begin with, let us briefly review the basic logic of the continuous orbifold approach of 
\cite{Gaberdiel:2011aa}. As was explained there, continuous compact groups (such as ${\rm U}(N)$)
behave in many respects like finite groups, and one may therefore believe that an orbifold  by 
a continuous compact group can be constructed essentially as in the familiar finite case. In particular,
the untwisted sector just consists of the invariant states of the original theory, while the twisted
sectors are labelled by the conjugacy classes of the orbifold group. Finally, in each such twisted
sector, only the states that are invariant with respect to the centraliser of the twist element 
survive.

For the case of ${\rm U}(N)$, the conjugacy classes are labelled by the elements
in the Cartan torus ${\rm U}(1)^N$ modulo the action of the Weyl group, i.e., the permutation
group $S_{N}$. Furthermore, the centraliser of a generic element of the Cartan torus is again
just the Cartan torus itself, i.e., the orbifold projection in the twisted sector will just guarantee
that the partition function is invariant under the $T$-transformation, $\tau\mapsto \tau+1$. 

\noindent Let us parametrise the elements of the Cartan torus by the diagonal matrices
\be\label{cartan}
\mathrm{diag}(e^{2\pi i\alpha_1},\ldots,e^{2\pi i\alpha_N}) \ , \qquad
-\tfrac{1}{2}<\alpha_i \leq \tfrac{1}{2} \quad (i=1,\ldots,N)\ . 
\ee
Since the Weyl group permutes these entries, the conjugacy classes (and thus the twisted sectors) 
can actually be labelled by
\be\label{twist}
 \alpha = [\alpha_1,\ldots,\alpha_N]\ ,
\ee
where now, in addition, $\alpha_i \leq \alpha_j$ for $i < j$.
In this section, we will argue that the ground state of the sector 
with twist $\alpha$ can be identified, in the limit $k\to\infty$, with the coset representative
\be\label{eq:gs0}
\Bigl(\Lambda_+(\alpha);\Lambda_-(\alpha),u(\alpha)\Bigr) \ ,
\ee
where $m \in \{0,\ldots,N\}$ is chosen such that
\be\label{alphacond}
  \alpha_i \leq 0 \text{ for } i \leq m\  \qquad\hbox{and} \qquad   \alpha_i \geq 0 \text{ for } i > m\ ,
\ee
and we define
\begin{align}\label{eq:gs_num}
\Lambda_+(\alpha)&= [k(\alpha_2-\alpha_1),
\ldots,k(\alpha_{m}-\alpha_{m-1}),-k\alpha_m,\\
&\qquad\qquad \qquad \qquad \qquad  k\alpha_{m+1},k(\alpha_{m+2}-\alpha_{m+1}),
\ldots, k(\alpha_{N}-\alpha_{N-1})]\ , \nonumber \\[8pt]
\Lambda_-(\alpha) &= [k (\alpha_2-\alpha_1),\ldots, k (\alpha_{N}-\alpha_{N-1})]\ ,\qquad \qquad \qquad
u(\alpha) = k \sum_{i=1}^N \alpha_i\ ,
\end{align}
where each entry of the weights is projected onto the integer part (and we also adjust $u(\alpha)$
correspondingly). These weights are then allowed at level $k$ since we have 
\be
\sum_{j=1}^{N+1}  \bigl[ \Lambda_+(\alpha) \bigr]_j = 
\sum_{j=1}^{N}  \bigl[ \Lambda_-(\alpha) \bigr]_j =  k (\alpha_N - \alpha_1 ) \leq k  \ .
\ee
One also easily checks that (\ref{eq:gs0}) satisfies the selection rule (\ref{sel}). 
Conversely,  for every coset primary $(\Lambda_+;\Lambda_-,u)$, we can write, 
after a suitable field redefinition if necessary, $\Lambda_+ \equiv \Lambda_+(\alpha)$ for some $\alpha$ 
of the form 
(\ref{twist}) with $-\tfrac{1}{2}<\alpha_i \leq \tfrac{1}{2}$ and $\alpha_i \leq \alpha_j$ for $i < j$;
indeed, the corresponding $\alpha$ may be taken to be 
\be
\alpha = \frac{1}{k} \,
\Biggl[-\sum_{i=1}^m \Lambda_i,-\sum_{i=2}^m \Lambda_i,\ldots,
-\Lambda_m,\Lambda_{m+1},\sum_{i=m+1}^{m+2}\Lambda_i,\ldots,
\sum_{i=m+1}^{N}\Lambda_i\Biggr]\ ,
\ee
where we choose $m$ such that 
\be
\sum_{i=1}^{m} \Lambda_i < \frac{k}{2} \ , \qquad \hbox{and} \qquad
\sum_{i=m+1}^{N} \Lambda_i \leq \frac{k}{2} \ . 
\ee
\smallskip

We will give three main pieces of evidence for this identification: we will show in  
section~\ref{sec:grounstate}
that the conformal dimension of the coset primary (\ref{eq:gs0}) agrees with the ground state 
energy of the $\alpha$-twisted state; we will confirm that the fermionic excitation spectrum of
the coset primary has the expected form (see section~\ref{sec:ex}); and we shall show 
in section~\ref{sec:BPS} that the twisted sector has BPS descendants precisely as suggested 
by the orbifold picture.

\subsection{Conformal dimension}\label{sec:grounstate}

In the $\alpha$-twisted sector the free fermions and bosons are simultaneously twisted
(as they transform in the same representation of ${\rm U}(N)$, see 
eq.~(\ref{transf}) above). As a consequence, the ground state 
energy of the $\alpha$-twisted sector should simply be
\be\label{htwist}
h(\alpha) = \frac{1}{2} \, \sum_{i=1}^{N} |\alpha_i| \ .
\ee
(For the convenience of the reader we have outlined the calculation of the 
twisted sector ground state energy in appendix~\ref{app:gs}, see in particular
eq.~\eqref{susytwist}.) We therefore need to show that the conformal dimension
of (\ref{eq:gs0}) agrees with (\ref{htwist}). 
\smallskip

In order to determine the conformal dimension of \eqref{eq:gs0},
we use \eqref{hdef} and note that the quadratic Casimir of a weight $\Lambda$
of $\mathfrak{su}(N)$ is given by
\be
C^{(N)}(\Lambda) = \sum_{i<j} \Lambda_i \Lambda_j \, \frac{i(N-j)}{N} 
+ \frac{1}{2}  \sum_{j} \Lambda_j^2 \, \frac{ j(N-j)}{N} 
+ \sum_{j} \Lambda_j \, \frac{j(N-j)}{2} \ . 
\ee
The key step of the computation is to calculate the difference of the Casimirs, which turns out to equal
\be\label{eq:gs_dC}
\Delta C = C^{(N+1)}\bigl(\Lambda_{+}(\alpha)\bigr)-C^{(N)}(\Lambda_{-}(\alpha))
= \frac{\left(k\sum_{i=1}^{N} \alpha_i\right)^2}{2N(N+1)}
+\frac{k}{2}\left(-\sum_{i=1}^m \alpha_i+\sum_{i=m+1}^N \alpha_i\right)\ .
\ee
It then follows that the conformal dimension is indeed given by
\begin{align} \label{eq:gs_h}
 h\Bigl(\Lambda_+(\alpha);\Lambda_-(\alpha),u(\alpha)\Bigr) & =  
\frac{\Delta C}{N+k+1}-\frac{u(\alpha)^2}{2N(N+1)(N+k+1)}  \\
&  =  \frac{k}{2(N+k+1)} \Bigl( -\sum_{i=1}^m \alpha_i+\sum_{i=m+1}^N \alpha_i\Bigr) 
\cong \frac{1}{2} \sum_{i=1}^N |\alpha_i|  \nonumber 
\end{align}
in the limit $k\to\infty$. Here we have used that the excitation number $n$ in \eqref{hdef} vanishes 
because the representation $\Lambda_-(\alpha)$ appears in the branching 
of $\Lambda_+(\alpha)$ from $\mathfrak{su}(N+1)$ to $\mathfrak{su}(N)$, as follows from
the discussion in appendix~\ref{app:branch}.

\noindent We should also mention that the $\mathrm{U}(1)$ charge of the coset primary equals
\be\label{eq:gs_q}
 q\Bigl(\Lambda_+(\alpha);\Lambda_{-}(\alpha),u(\alpha)\Bigr)=
\frac{u(\alpha)}{N+k+1} \cong \sum_{i=1}^N \alpha_i\ ,
\ee
which also agrees with what one expects based on the twisted sector analysis. Note that 
the ground state is a chiral primary 
if all twists are positive, and an anti-chiral primary if all twists
are negative; we shall come back to a more detailed analysis of the BPS states in the twisted
sectors in section~\ref{sec:BPS}.

\subsection{The fermionic excitation spectrum}\label{sec:ex}

We can test the above correspondence further by calculating the 
actual excitation spectrum of the fermions in the twisted sector. Recall that 
the free fermions correspond to the coset primaries $(0;{\rm f},-(N+1))$ and 
$(0;\bar{\rm f},(N+1))$, respectively. We can therefore determine the `twist' of these
fermions by evaluating the change in conformal dimension upon fusion
with these fields. As a by-product of this analysis we will also be able to show that the above
coset primaries are indeed ground states.

More specifically, suppose that $(\Lambda_+;\Lambda_-,u)$ is the (ground) state
of a twisted sector. Then we consider the fusion products
\be\label{fermdesc}
(\Lambda_+;\Lambda_-,u) \otimes \bigl(0;{\rm f},-(N+1) \bigr) 
= \bigoplus_{l=0}^{N-1} \bigl(\Lambda_+;\Lambda_-^{-(l)} , u - (N+1) \bigr) \ , 
\ee
where $\Lambda^{-(l)}$ with $l=0,\ldots, N-1$ denotes the $N$ different weights
that appear in the tensor product $\Lambda\otimes {\rm f}$. Similarly we define 
\be\label{fermbardesc}
(\Lambda_+;\Lambda_-,u) \otimes \bigl(0;\bar{\rm f},(N+1) \bigr) 
= \bigoplus_{l=0}^{N-1} \bigl(\Lambda_+;\Lambda_-^{+(l)} , u + (N+1) \bigr) \ , 
\ee
where $\Lambda^{+(l)}$  labels the weights that appear in $\Lambda\otimes \bar{\rm f}$;
a closed formula for both cases is given by 
\be\label{tensorl}
\Lambda^{\epsilon(l)}_j = \left\{
\begin{array}{ll}
\Lambda_j + \epsilon \qquad & j=l \\
\Lambda_j - \epsilon \qquad & j=l+1 \\
\Lambda_j \qquad & \hbox{otherwise.}
\end{array}
\right. 
\ee
Here $\epsilon = \pm$, and we have assumed that all $\Lambda_j\neq 0$ so that all 
$N$ fusion channels $\Lambda^{\epsilon (l)}$ are indeed allowed. (We will comment on
the situation when this is not the case at the end of this subsection.) 

Now the `twist' of the  fermionic excitations of the twisted sector state
$(\Lambda_+;\Lambda_-,u)$ can be determined by calculating the difference of 
conformal dimension of the coset primaries that appear in (\ref{fermdesc}) and 
(\ref{fermbardesc}), relative to the original state. Indeed, generically, there will be 
$N$ different such twists, corresponding to the $N$ different fusion channels in (\ref{tensorl}), and this
ties in with the fact that there are $N$ fundamental fermions (as well as their conjugates). 
One cross-check of our analysis will be that the twists of the fermions and 
their conjugates will be opposite, and this will indeed turn out to be the case.

In order to calculate this difference of conformal dimension we note that it
follows from (\ref{hdef}) that 
\begin{align}
\delta h^{(l)} & \equiv  h \Bigl(\Lambda_+;\Lambda_-^{\epsilon(l)},u+ \epsilon (N+1) \Bigr)  - 
h (\Lambda_+;\Lambda_{-},u) \nonumber \\
& =  \frac{1}{N+k+1}\, \Bigl( 
C^{(N)}(\Lambda_{-}) - C^{(N)}(\Lambda^{\epsilon(l)}_{-}) \Bigr) \nonumber \\
&  \qquad 
{}-  \frac{1}{2N(N+1)(N+k+1)} \Bigl( 2 \epsilon u (N+1) + (N+1)^2 \Bigr) + n \ . 
\end{align}
The difference of Casimir operators turns out to equal
\begin{align}
\delta C^{(l)} & =  C^{(N)}(\Lambda_-) - C^{(N)}(\Lambda^{\epsilon(l)}_-)  \nonumber \\
& =  - \frac{\epsilon}{N} \sum_{i=1}^{N-1} \, i\, \Lambda_i + \epsilon \sum_{j=l+1}^{N-1} \Lambda_j  
+ \frac{1}{2N} \bigl( \epsilon N^2 - 2 l \epsilon N - \epsilon N + 1 - N \bigr) \ ,
\end{align}
where $\Lambda_j$ are the Dynkin labels of $\Lambda_-$. 
Thus we find that 
\begin{align}
\delta h^{(l)} & =  
n + \frac{1}{N+k+1}  \Bigl[
- \frac{\epsilon}{N}  \Bigl( \sum_{i=1}^{N-1} \, i\, \Lambda_i +  u \Bigr) 
+ \epsilon \sum_{j=l+1}^{N-1} \Lambda_j  \Bigr] \nonumber \\
& \quad {} + \frac{1}{2 (N+k+1)} (\epsilon N - 2 l \epsilon - (2+\epsilon)) \ . 
\end{align}
In the limit $k\rightarrow \infty$, the second line can be ignored (since none of the terms in 
the numerator can depend on $k$), and hence we get approximately
\be\label{shift}
\delta h^{(l)} \cong n + \frac{\epsilon}{N+k+1} \,
\Bigl[\, \sum_{j=l+1}^{N-1} \Lambda_j 
-  \frac{1}{N}  \Bigl( \sum_{i=1}^{N-1} \, i\, \Lambda_i +  u \Bigr) \Bigr] \ .
\ee
Applying this formula to the state \eqref{eq:gs0} and using (\ref{eq:gs_num}) yields then
\be\label{shift_gs}
\delta h^{(l)} \cong n - \epsilon\, \alpha_{l+1}\ ,
\ee
where $\alpha_{l+1}$ denotes the different components of the twist in (\ref{twist}). 
For the free fermions,\footnote{Technically, this means we have to consider the so-called 
`even' fusion of the associated coset fields, see \cite{Mussardo:1988av,Mussardo:1988ck},
as well as \cite{Gaberdiel:1996kf}. In order to analyse the bosonic descendants (that sit in
the same ${\cal N}=2$ representation), we then have to consider the `odd' fusion rules.}
the selection rule of the $\mathfrak{so}(2N)_1$ factor implies 
that $n=\frac{1}{2}$. Thus, the excitations of the fermions are shifted away from
the untwisted NS value $\delta h= \frac{1}{2}$ by the twist $\alpha_{l+1}$. Furthermore, this twist 
is opposite for the fermions and the anti-fermions, i.e., it is proportional to $\epsilon$. 
This then agrees precisely with what should be the case for the $\alpha$-twisted sector.
\smallskip

It is worth stressing that the derivation of \eqref{shift} was completely general, and did, in particular,
not assume any specific properties of the state $(\Lambda_+;\Lambda_-,u)$. Thus 
we can use it to read off the twist of {\em any} coset state, which therefore equals
\be\label{eq:twist_gen}
\alpha_j \cong -\frac{1}{N+k+1} \,
\Bigl[\, \sum_{i=j}^{N-1} \Lambda_j -  
\frac{1}{N}  \Bigl( \sum_{i=1}^{N-1} \, i\, \Lambda_i +  u \Bigr) \Bigr]\ ,
\ee
where the $\Lambda_j$ are, as before, the Dynkin labels of $\Lambda_-$. 
Note that finite excitations only change the $\Lambda_i$ and $u$
by a finite amount, which can be neglected in the limit  $k\rightarrow \infty$. We therefore conclude that 
finitely excited states live in the same twisted sector as the corresponding ground state.
Again, this is what should be the case for the $\alpha$-twisted sector.

Finally, we comment on the special situation for which some of the $\Lambda_j=0$. 
In that case, there are actually fewer fermionic excitations since some of the $l$ in 
(\ref{tensorl}) are not allowed. This phenomenon also has a very natural interpretation from the 
continuous orbifold perspective: because of eq.~(\ref{eq:gs_num}),
$\Lambda_j=0$ implies that $\alpha_{j+1} = \alpha_j$. Then the centraliser of the corresponding
element of the Cartan torus (\ref{cartan}) is {\em bigger} than just the Cartan torus itself, since
it includes, in particular, the ${\rm SU}(2)$ subgroup that rotates the two twists $\alpha_j$ 
and $\alpha_{j+1}$ into one another. This means that actually fewer fermionic excitations
survive the orbifold projection in the twisted sector, in perfect agreement with the fact that we also have
fewer coset descendants. The analysis works similarly if more than one $\Lambda_j=0$, etc.

It remains to show that the coset states \eqref{eq:gs0} actually correspond to the {\em ground
states} of the $\alpha$-twisted sector. For the fermionic excitations with $n=\frac{1}{2}$ this is
obvious from the above (given that, by construction, each $|\alpha_j|\leq \frac{1}{2}$).  The argument
for the bosonic descendants (for which $n=0$ is possible) requires more work and is spelled out 
in appendix~\ref{gs}.

\subsection{BPS descendants}\label{sec:BPS}

Finally, we want to analyse the BPS descendants of the twisted sector ground states.
For the case with ${\cal N}=4$ superconformal symmetry, it is well known from the 
analysis of the symmetric orbifold, see e.g., \cite{Lunin:2001pw},  that 
each twisted sector of the symmetric orbifold contains a BPS descendant that is obtained from the 
twisted sector ground state upon applying all fermionic generators whose mode number is less 
than $1/2$.  For the case at hand, i.e., the situation with ${\cal N}=2$ superconformal symmetry, 
we expect that each twisted sector should contain {\em two} BPS states, one chiral primary that is obtained 
by applying all $q=+1$ fermionic modes whose mode number is less than $1/2$ to the twisted sector 
ground state; and one anti-chiral primary that is obtained by applying all $q=-1$ fermionic modes whose 
mode number is less than $1/2$. Actually, as we shall see, 
this expectation is borne out; quite surprisingly, the relevant chiral and anti-chiral states remain
BPS even at finite $N$ and $k$. 

To be more specific, let us consider the twisted sector ground state 
defined in eq.~(\ref{eq:gs0}).
In order to obtain the chiral primary descendant we have to apply the fermionic
modes associated to $(0;{\rm f},-(N+1))$ whose mode numbers are less than $1/2$. 
Thus we should consider the descendant where we add a box to each of the first $m$ rows,
i.e., the coset primary 
\be
\Bigl( \Lambda_+(\alpha);\Lambda_-(\alpha)^{({\rm BPS})},u(\alpha)^{({\rm BPS})} \Bigr) \ , 
\ee
where, for $m\geq 1$,
\be
\Lambda_-(\alpha)^{({\rm BPS})} =
\Bigl[ k(\alpha_2-\alpha_1),\ldots,k (\alpha_{m+1}-\alpha_m)+1,\ldots, k(\alpha_N-\alpha_{N-1})\Bigr]\ , 
\ee
and 
\be
u(\alpha)^{({\rm BPS})} = k\sum_{i=1}^N \alpha_i - m (N+1) \ . 
\ee
We now claim that this defines a chiral primary operator, even for finite $N$ and $k$.\footnote{One way
to see this is to note that, up to a field identification, this coset primary satisfies
$\Lambda_- = P \Lambda_+$, where $P$ is the restriction to the first $N-1$ Dynkin labels. We thank
Stefan Fredenhagen for pointing this out to us.}
Similarly, 
the anti-chiral primary is obtained by applying the fermionic modes associated to $(0;\bar{\rm f},(N+1))$
whose mode numbers are less than $1/2$, i.e., by removing a box in each of the rows $m+1,\ldots, N$. The 
corresponding anti-chiral primary is then 
\be
\Bigl( \Lambda_+(\alpha);\Lambda_-(\alpha)^{(\overline{\rm BPS})},u(\alpha)^{(\overline{\rm BPS})} \Bigr) \ , 
\ee
where, for $m<N$,
\be
\Lambda_-(\alpha)^{(\overline{\rm BPS})} =
\Bigl[ k(\alpha_2-\alpha_1),\ldots, k(\alpha_{m+1}-\alpha_m)+1, \ldots,k (\alpha_N-\alpha_{N-1})\Bigr]
= \Lambda_-(\alpha)^{({\rm BPS})} \ , 
\ee
but now 
\be
u(\alpha)^{(\overline{\rm BPS})} = k\sum_{i=1}^N \alpha_i + (N-m) (N+1) \ . 
\ee
Note that both states satisfy the selection rule \eqref{sel} because 
\be
\abs{\Lambda_+(\alpha)} =-k\sum_{i=1}^N \alpha_i +(N+1)k\alpha_N
\ee
and 
\be
\abs{\Lambda_-(\alpha)^{({\rm BPS})} }  = \abs{\Lambda_-(\alpha)^{(\overline{\rm BPS})} } 
= -k\sum_{i=1}^N \alpha_i +N k\alpha_N+m\ .
\ee
To show that these states are indeed chiral and anti-chiral primaries, 
we again first compute the difference of the Casimirs; using the result from \eqref{eq:gs_dC} we 
obtain
\begin{align}
\Delta C & =  C^{(N+1)}\bigl(\Lambda_+(\alpha)\bigr) - C^{(N)} \bigl(\Lambda_-(\alpha)^{({\rm BPS})}\bigr) \nonumber \\
&=  C^{(N+1)}\bigl(\Lambda_+(\alpha)\bigr)- C^{(N)}\bigl(\Lambda_-(\alpha)\bigr)  
+k\sum_{i=1}^m \alpha_i-\frac{mk}{N}\sum_{i=1}^N\alpha_i-
\frac{N+1}{2N}m(N-m) \nonumber \\
&= \frac{\bigl[u(\alpha)^{({\rm BPS})}\bigr]^2}{2N(N+1)}
+\frac{1}{2} u(\alpha)^{({\rm BPS})} \ .
\end{align}
Eqs.~\eqref{hdef} and \eqref{qdef} then directly lead to
\begin{align}
h\Bigl(\Lambda_+(\alpha);\Lambda_-(\alpha)^{({\rm BPS})},u(\alpha)^{({\rm BPS})}\Bigr)
& =  \frac{u(\alpha)^{({\rm BPS})}}{2(N+k+1)}+\frac{m}{2} \\[2pt]
& =  \frac{1}{2}\, q\Bigl(\Lambda_+(\alpha);\Lambda_-(\alpha)^{({\rm BPS})},u(\alpha)^{({\rm BPS})}\Bigr)\ ,
 \nonumber
\end{align}
so these states are indeed chiral primary.
Similarly, using 
\be
u(\alpha)^{({\rm BPS})} =u(\alpha)^{(\overline{\rm BPS})} -N(N+1)
\ee
 we compute
\begin{align}
h\Bigl(\Lambda_+(\alpha);\Lambda_-(\alpha)^{(\overline{\rm BPS})},u(\alpha)^{(\overline{\rm BPS})}\Bigr)
& = -\frac{u(\alpha)^{(\overline{\rm BPS})}} {2(N+k+1)}+\frac{N-m}{2}  \\[2pt]
& = -\frac{1}{2}\, q\Bigl(\Lambda_+(\alpha);\Lambda_-(\alpha)^{(\overline{\rm BPS})},
u(\alpha)^{(\overline{\rm BPS})}\Bigr)\ , \nonumber
\end{align}
and thus these states are anti-chiral primary as claimed.

Note that for $m=0$, all twists are non-negative, and so by \eqref{eq:gs_h} and
\eqref{eq:gs_q} already the ground state is chiral primary. Similarly,
the ground state with $m=N$ is anti-chiral primary since all twists are non-positive.

\section{Conclusions and outlook}\label{sec:5}

In this paper we have collected evidence for the assertion
that the $\mathcal{N}=2\,$ $\mathrm{SU}(N)$ Kazama-Suzuki models that occur in the duality
with the higher spin theory on AdS$_3$ can be described,
in the $k\to\infty$ limit, by a $\mathrm{U}(N)$ orbifold of $2N$ free bosons and fermions.
In particular, the subsector of the coset theory consisting of the states of the form 
$(0;\Lambda,u)$ --- these are dual to the  excitations of one complex scalar multiplet of the higher spin theory --- 
corresponds to the untwisted sector of this orbifold, as follows from the comparison of the
partition functions.  We have also identified
the twisted sector ground states from the coset perspective, and shown
that their conformal dimension, their excitation spectrum and their BPS 
descendants match the orbifold prediction. In particular, the BPS states
are generated from the ground states by exciting them with all fermions
or antifermions whose twist has the same sign.

Our analysis was motivated by the duality  \cite{Gaberdiel:2013vva} relating the family of 
Wolf space cosets with large $\mathcal{N}=4$ superconformal symmetry to the 
${\cal N}=4$ superconformal higher spin theory on AdS$_3$. In this case, the $k\rightarrow \infty$
limit corresponds to the situation where the radius of one of the two ${\rm S}^3$'s in 
${\rm AdS}_3 \times {\rm S}^3 \times {\rm S}^3 \times {\rm S}^1$ becomes infinite,
and one may hope to make contact with string theory on 
${\rm AdS}_3 \times {\rm S}^3 \times \mathbb{T}^4$; this 
will be explored in more detail elsewhere~\cite{GGprep}.

\section*{Acknowledgments}
We  thank Constantin Candu, Stefan Fredenhagen and Rajesh Gopakumar for very helpful discussions
and comments, and Stefan Fredenhagen for drawing our attention to \cite{Restuccia:2013tba}. 
This work is partially supported by a grant of the Swiss National Science Foundation.
MRG thanks the Weizmann Institute, Rehovot, Chulalongkorn University, Bangkok, 
as well as the University of North Carolina in Chapel Hill for hospitality during various
stages of this work.

\appendix
\section{Twisted sector ground state energies}
\label{app:gs}
In this appendix we collect together some formulae for the ground state energies of twisted fermions and bosons.

\subsection{Complex free fermions}

We begin with the case of free fermions
twisted by $\alpha$ with $-\frac{1}{2}\leq\alpha\leq \frac{1}{2}$.
Let us consider a pair of complex fermions that pick up eigenvalues $e^{\pm 2\pi i \alpha}$
under the twist. The relevant twining character,
i.e.\ the character with the insertion of the 
eigenvalues $e^{\pm 2\pi i \alpha}$, equals then in the NS-sector
\be
\chi_\alpha(\tau) = \frac{\vartheta_3(\tau,\alpha)}{\eta(\tau)} \ , 
\ee
where we use the definitions 
\begin{align}
\eta(\tau) & = q^{\frac{1}{24}} \, \prod_{n=1}^{\infty} (1-q^n)\ , \\[2pt]
\vartheta_3(\tau,z) & = \prod_{n=1}^{\infty} (1-q^n)\,  (1+ y q^{n-1/2})\, (1+y^{-1} q^{n-1/2})  \ , 
\end{align}
as well as $q=e^{2\pi i \tau}$ and $y=e^{2\pi i z}$. To obtain the ground state energy of the twisted sector
we perform an $S$-modular transformation, using the transformation rules
\begin{align}
\eta(-\tfrac{1}{\tau}) & = (-i \tau)^{1/2}\, \eta(\tau) \\[2pt]
\vartheta_3(-\tfrac{1}{\tau},\tfrac{z}{\tau}) & =  (-i \tau)^{1/2}\, e^{i \pi z^2/ \tau}\, \vartheta_3(\tau,z)  \ ,
\end{align}
to obtain for the $\alpha$-twisted partition function 
\begin{align}
\chi_\alpha(-\tfrac{1}{\tau}) & = e^{i \pi \alpha^2 \tau} \, \frac{\vartheta_3(\tau,\tau\alpha)}{\eta(\tau)}  \\[2pt]
& = q^{-\tfrac{1}{24}}\, e^{i \pi \alpha^2 \tau} \, \prod_{n=1}^{\infty} (1 + e^{2\pi i \tau \alpha} q^{n-1/2}) \, 
(1 + e^{-2\pi i \tau \alpha} q^{n-1/2})  \ .
\end{align}
Thus the ground state energy of the $\alpha$-twisted sector equals
\be
\Delta h_{\rm fer} = \frac{1}{2} \alpha^2 \ .
\ee

\subsection{Complex free bosons and susy case}

The analysis for a pair of complex bosons is essentially identical. Now the relevant twining character equals
\be
\chi_\alpha(\tau) = - 2 \sin(\pi \alpha)\, \frac{\eta(\tau)} {\vartheta_1(\tau,\alpha)}\ , 
\ee
where $\vartheta_1(\tau,z)$ is defined by
\be
\vartheta_1(\tau,z)  =  - 2 q^{1/8} \sin(\pi z)\,  \prod_{n=1}^{\infty} (1-q^n)\,  (1- y q^{n})\, (1-y^{-1} q^{n})  \ .
\ee
The modular transformation behaviour of $\vartheta_1(\tau,z)$ is 
\be
\vartheta_1(-\tfrac{1}{\tau},\tfrac{z}{\tau}) = 
- i (i\tau)^{1/2} e^{i \pi z^2/ \tau}\, \vartheta_1(\tau,z) \ ,
\ee
and hence the twisted character equals
\be
\chi_\alpha(-\tfrac{1}{\tau}) = i \frac{\sin(\pi \alpha)}{\sin(\pi \tau \alpha)}\, e^{-i \pi \alpha^2 \tau}\,  q^{-\frac{2}{24}} \, 
\prod_{n=1}^{\infty} \frac{1}{(1- e^{2\pi i \alpha\tau} q^{n})\, (1-e^{-2\pi i \alpha\tau} q^{n})} \ . 
\ee
For $-\frac{1}{2}\leq \alpha\leq \frac{1}{2}$ 
we read off from the leading $q\rightarrow 0$ behaviour that
\be
\Delta h_{\rm bos} = \frac{1}{2} |\alpha| - \frac{1}{2} \alpha^2\ .
\ee

Note that for a supersymmetric theory, i.e., for a theory where both bosons and fermions 
are twisted by the same amount, the total ground state energy is then
\be\label{susytwist}
\Delta h_{\rm tot}= \Delta h_{\rm bos} + \Delta h_{\rm fer} = \frac{|\alpha|}{2} \ , 
\ee
which is indeed linear in $|\alpha|$. 

\section{Branching rules}\label{app:branch}

In this appendix we explain the branching
rules of $\mathfrak{su}(N+1)\supset \mathfrak{su}(N)$.
They were first derived by Weyl \cite{Weyl:1931} 
in terms of $\mathfrak{u}(N)$ tensors 
(see, e.g., \cite{Whippman:1965} and \cite{King:1975vf} for more modern
and general treatments).

\smallskip
Let $\Lambda=[\Lambda_1,\ldots,\Lambda_N]$ be a 
highest weight of $\mathfrak{su}(N+1)$. 
The procedure can be divided into three steps:
\begin{enumerate}
\item
Interpret $\Lambda$ as a highest weight of $\mathfrak{u}(N+1)$ 
rather than $\mathfrak{su}(N+1)$.
\item
Let $r_i$ denote the number
of boxes in the $i$\textsuperscript{th} row of the Young diagram
associated with $\Lambda$, 
\be
r_i = \sum_{j=i}^N \Lambda_j\ .
\ee
Then under the branching $\mathfrak{u}(N+1)\supset \mathfrak{u}(N)$
we have the decomposition

\be
\Lambda \to \bigoplus_{\tilde\Lambda} \tilde\Lambda\ ,
\ee
where $\tilde\Lambda=[\tilde\Lambda_1,\ldots,\tilde\Lambda_N]$
are highest weights of $\mathfrak{u}(N)$ whose rows $\tilde r_i$ satisfy
\be\label{eq:Weyl_br}
 r_1 \geq \tilde r_1 \geq r_2 \geq \tilde r_2 \geq \cdots\geq \tilde r_N \geq 0\ ,
\ee
each $\tilde\Lambda$ appearing once.
\item
In the end, each $\tilde\Lambda$ has to be restricted to $\mathfrak{su}(N)$
by removing the last Dynkin label.
\end{enumerate}
Equation~\eqref{eq:Weyl_br} means that from each row $i=1,\ldots,N$,
any number $a_i=0,\ldots,\Lambda_i$ of boxes may be removed, such that
the new number of boxes in the $i$\textsuperscript{th} row becomes
\be
\tilde r_i =r_i-a_i\ .
\ee
So the weights $\tilde\Lambda$ are labelled by
the  vectors $\mathbf{a}=(a_1,\ldots,a_N)$ and we write
$\Lambda(\mathbf{a})$ for the restriction
to $\mathfrak{su}(N)$ of the $\tilde\Lambda$ 
labelled by $\mathbf{a}$. The Dynkin labels
of $\Lambda(\mathbf{a})$ are given by
\be
\Lambda(\mathbf{a})_i = \tilde r_i - \tilde r_{i+1} = \Lambda_i - a_i + a_{i+1}
\ee
for $i=1,\ldots,N-1$, and thus the branching rules may be written as 
\be
\Lambda \to \bigoplus_{\bf a} \Lambda({\bf a}) \ .
\ee

\section{The ground state analysis}\label{gs}

In this appendix we shall show that the coset states (\ref{eq:gs0}) actually define twisted sector
ground states. In particular, we need to show that $\delta h^{(l)}$ in (\ref{shift_gs}) is non-negative for all
$l=0,\ldots,N-1$. Because the individual twists satisfy $|\alpha_i|\leq \frac{1}{2}$, only the representations 
with $n=0$ have a chance of  lowering the conformal dimension of 
the original state. The $\mathfrak{so}(2N)_1$ selection rule implies that $n=\tfrac{1}{2}$
for the actual fermionic excitations, but $n=0$ can arise for the bosonic excitations (that come from 
the same multiplets). Thus we need to analyse (i) whether $n=0$ is allowed in the fusion with 
$(0;{\rm f},-(N+1))$ or $(0;\bar{\rm f},(N+1))$; and (ii) if so, whether the relevant term in (\ref{shift}) is 
then positive. 
\smallskip

The condition that $n=0$ is possible simply means that $\Lambda_-(\alpha)^{\epsilon (l)}$
is contained in $\Lambda_+(\alpha)$ under the branching rules of $\mathfrak{su}(N+1)\supset \mathfrak{su}(N)$. 
In the notation of appendix~\ref{app:branch} the original coset state (\ref{eq:gs0}) corresponds to the choice
$\Lambda_+(\alpha) \equiv \Lambda$, and $\Lambda_- (\alpha)\equiv \Lambda({\bf a})$ with 
\be\label{aform}
{\bf a} = {\bf a}^{(m)} =  (0,\ldots,0,\Lambda_{m+1},\ldots, \Lambda_N) \ . 
\ee
Furthermore, generically the fusion with $(0;{\rm f},-(N+1))$ or $(0;\bar{\rm f},(N+1))$ leads to 
\be\label{ashift}
\Lambda(\mathbf{a})^{\epsilon(l)} = \Lambda(\mathbf{a}')\,, \quad
\hbox{where} \quad
a'_j = \left\{ \begin{array}{ll}
a_j \quad & j\neq l+1 \\
a_{l+1} + \epsilon \quad & j=l+1\ .
\end{array}
\right.
\ee
However, this representation only appears in the above branching rules of
the same $\Lambda_+(\alpha)\equiv \Lambda$ if all $a'_j$ satisfy 
$0\leq a'_j \leq \Lambda_j$. Thus we see that $n=0$ is only allowed if 
\begin{align}
\hbox{for $\epsilon=+$} & \qquad   a_{l+1} < \Lambda_{l+1} \  \ \ \hbox{i.e.,} \  \ 
l \leq m \nonumber \\[4pt]
\hbox{for $\epsilon=-$} & \qquad   0< a_{l+1}  \ \ \ \hbox{i.e.,} \ \
l \geq m+1 \ . 
\end{align}
(We are assuming here, for simplicity, that all $\Lambda_j\neq 0$.) But for these
values of $\epsilon$ and $l$,  it then follows from  (\ref{alphacond}) 
that $-\epsilon \, \alpha_{l+1} \geq  0$. This therefore shows that  $\delta h^{(l)}$ in (\ref{shift_gs}) is 
indeed non-negative.

\subsection{Other potential twisted sector ground states}

It is also not hard to show that among the `light states', i.e., those that have $n=0$, 
the only twisted sector ground states are in fact those described in (\ref{eq:gs0}).  The most
general light states are of the form 
\be
 \label{eq:gs1}
\Bigl(\Lambda;\Lambda({\bf a}),-|\Lambda| + (N+1) \sum_{j=1}^{N} a_j \Bigr) \ , \qquad
|\Lambda| = \sum_{j=1}^{N}\, j \,  \Lambda_j \ , 
\ee
where ${\bf a}=(a_1,\ldots,a_N)$, and the $a_i$ take the values $a_i=0,\ldots, \Lambda_i$, $i=1,\ldots,N$.
We want to show that  among these states, the only ones that are twisted sector ground states, i.e., annihilated 
by all positive  fermionic and bosonic modes, are those for which ${\bf a}$ is of the form (\ref{aform}). In 
order to analyse this issue, we determine the analogue of 
(\ref{shift}), which now takes the form
\be\label{shift_gen}
\delta h^{(l)} \cong n +\frac{\epsilon}{N+k+1} \,
\Biggl(\, (\Lambda_{l+1}-a_{l+1})+\sum_{i=l+2}^{N} \Lambda_i - A \Biggr)\,,
\qquad A=\sum_{i=1}^{N} a_i \ .
\ee
Using (\ref{ashift}), we have again that $n=0$ is only allowed for $\epsilon=+$ if $a_{l+1} < \Lambda_{l+1}$, and for 
$\epsilon=-$ if $a_{l+1}>0$ --- otherwise the representation $\Lambda({\bf a}')$ 
does not appear in the branching rules of 
$\mathfrak{su}(N+1) \supset \mathfrak{su}(N)$. 
It follows that if $0< a_j < \Lambda_j$, both values $\epsilon=\pm$ allow for
$n=0$ and thus one of the two $\delta h^{(l)}$ will be negative. So for a ground state,
each $a_j$ is either $a_j=0$ or $a_j=\Lambda_j$.

As a last step, we show that in fact $\mathbf{a}=\mathbf{a}^{(m)}$ for some
$m=0,\ldots,N$. Requiring \eqref{shift_gen} to be non-negative for all $l$, we obtain
the inequalities
\be
\hbox{if $a_{l+1} = 0$}\ : \qquad A \leq \sum_{j=l+1}^{N} \Lambda_j 
\ee
(recall that for $a_{l+1}=0$, $n=0$ occurs for $\epsilon=+$) and
\be
\hbox{if $a_{l+1} = \Lambda_{l+1}$}\ : \qquad \sum_{j=l+2}^{N} \Lambda_j \leq A 
\ee
(since for $a_{l+1}=\Lambda_{l+1}$, $n=0$ occurs for $\epsilon=-$). 
\smallskip

The sequence of partial sums $P_r = \sum_{j=r}^{N} \Lambda_j$ is strictly decreasing, whereas
$A$ takes the same value in all of these inequalities. This implies that the $a_j$ 
have to be chosen in such a way that 
$\mathbf{a}=(0,\ldots,0,\Lambda_{m+1},\ldots,\Lambda_N)=\mathbf{a}^{(m)}$. This completes the proof.

\end{document}